\documentclass[11pt]{article}
\usepackage{moriond,epsfig}
\usepackage[latin1]{inputenc}
\usepackage[OT1]{fontenc}
\usepackage{amssymb}
\usepackage{amsmath}

\begin{document}


\title{SCALAR PERTURBATIONS IN DGP BRANEWORLD COSMOLOGY}

\author{ANTONIO CARDOSO~\footnote{Based on the work~\cite{Cardoso:2007xc} done in collaboration with Kazuya Koyama, Sanjeev S. Seahra and Fabio P. Silva.}}

\address{Institute of Cosmology \& Gravitation, University of
Portsmouth, Portsmouth~PO1~2EG, UK}

\maketitle

\abstracts{We solve for the behaviour of cosmological perturbations
in the Dvali-Gabadadze-Porrati (DGP) braneworld model using a new
numerical method. Unlike some other approaches in the literature,
our method uses no approximations other than linear theory and is
valid on large scales. We examine the behaviour of late-universe
density perturbations for both the self-accelerating and normal
branches of DGP cosmology.}

\section{Introduction}

The Dvali-Gabadadze-Porrati (DGP)~\cite{Dvali:2000rv,Dvali:2000hr}
model postulates that we live in a 4-dimensional hypersurface in a
5-dimensional Minkowski bulk. General Relativity (GR) is recovered
at small scales (smaller than the crossover scale $r_{\rm{c}}$) due
to the inclusion of an induced gravity term in the action.

This model has two distinct classes of cosmological
solutions.~\cite{Deffayet:2000uy} One of them exhibits accelerated
expansion at late times without the need to include any exotic
cosmological fluids, such as dark energy, or any brane tension that
acts as an effective 4-dimensional cosmological constant. Hence,
this branch of solutions is called ``self-accelerating''. To explain
the observed acceleration we require $r_{\rm{c}} \sim H_0^{-1}$,
where $H_0$ is the current value of the Hubble parameter. It is
expected that structure formation will help to distinguish the
self-accelerating DGP universe from dark energy models based on
4-dimensional GR. This is because the growth of cosmological
perturbations is very sensitive to the existence of an extra
dimension. A full 5-dimensional treatment is required to model these
perturbations, which is why obtaining observational predictions for
the behaviour of fluctuations in the DGP model is technically
challenging.

Several authors have considered the problem of the dynamics of
perturbations in the DGP model, but they have all relied on some
sort of approximation or simplifying \emph{ansatz}. One example of
this is the quasi-static (QS)~\cite{Koyama:2005kd} approximation
scheme, which solves the perturbative equations of motion by
focussing on the extreme subhorizon regime.

In this paper we present a complete numerical analysis of the
evolution of scalar perturbations in the DGP model. Mathematically,
the problem involves the solution of a partial differential equation
in the bulk coupled to an ordinary differential equation on the
brane. A numerical method for dealing with such systems has
previously been developed for cosmological perturbations in the
Randall-Sundrum (RS) model.~\cite{Cardoso:2006nh,Cardoso:2007zh}
However, the DGP problem is more complicated than the RS case due to
a non-local boundary condition on the bulk field. Hence, the
algorithm used in this paper represents a significant generalization
of the one used in the previous case.

As alluded to above, there is another ``normal'' branch of solutions
in the DGP model. We cannot explain the late time accelerated
expansion of the Universe using the normal branch without including
an effective cosmological constant induced by the brane tension
$\sigma$. However, by allowing a non-zero $\sigma$, normal branch
solutions can mimic dark energy models with the equation of state
$w$ smaller than $-1$.~\cite{Sahni:2002dx,Lue:2004za} Unlike
4-dimensional models that realize $w < -1$ by the introduction of a
phantom field, the normal branch of DGP cosmology is
ghost-free.~\cite{Charmousis:2006pn} This unique feature is what
motivates us to numerically study the perturbations of the normal
branch of DGP cosmology.

\section{Background solution}\label{sec:background}

In the DGP model the brane dynamics is governed by
\begin{align}\label{eq:background dynamics}
    H & = \frac{\dot{a}}{a} = \frac{1}{2r_{\rm{c}}} \left[ \epsilon + \sqrt{1 + \frac{4}{3}\kappa_4^2 r_{\rm{c}}^2
    (\rho+\sigma)}\right], \\
    \frac{d\rho}{dt} & = -3(1+w)\rho H,
\end{align}
where $a(t)$ is the scale factor of the brane universe, normalized
to unity today, a dot denotes the derivative with respect to the
proper time along the brane, $t$, and
$r_{\rm{c}}=\kappa_5^2/2\kappa_4^2$. We assume that the
stress-energy tensor of the brane matter is of the perfect fluid
form. Note that, as in GR, the stress-energy tensor is conserved.
The $\epsilon = \pm 1$ parameter reflects the fact that when we
impose the $\mathcal{Z}_2$ symmetry across the brane, we have two
choices for the half of the bulk manifold we discard. When $\epsilon
= +1$, we have that $Hr_{\rm{c}} \approx 1$ when the density of
brane matter is small, $|\rho + \sigma| \ll \kappa_4^{-2}
r_{\rm{c}}^{-2}$. This implies a late-time accelerating universe,
which is why the $\epsilon = +1$ case is called the
self-accelerating branch and the $\epsilon = -1$ case is called the
normal branch.

\section{Master equations governing perturbations}\label{sec:master}

In this section we describe the formulae governing scalar
perturbations in the late-time matter dominated universe. We take
the matter content of the brane to be a dust fluid (i.e., cold dark
matter), $w = 0$, therefore we have $\rho \propto a^{-3}$.

It can be shown that the dynamics of all the scalar perturbations of
the bulk geometry can be derived from a single scalar bulk degree of
freedom.~\cite{Mukohyama:2000ui} After Fourier decomposition, we
find that the mode amplitude $\Omega = \Omega(u,v)$ of this master
field obeys
\begin{equation}\label{eq:scalar master eqn}
    0 = \frac{\partial^2 \Omega}{\partial u \partial v} - \frac{3}{2v} \frac{\partial\Omega}{\partial
    u} + \frac{k^2 r_{\rm{c}}^2}{4v^2} \Omega,
\end{equation}
where $u$ and $v$ are dimensionless null coordinates.

We also find the following boundary condition for $\Omega$:
\begin{equation}\label{eq:scalar boundary condition}
    (\partial_y \Omega)_{\rm{b}} = - \frac{\epsilon \gamma_1}{2H}
    \ddot\Omega_{\rm{b}} +\frac{9\epsilon \gamma_3}{4} \dot\Omega_{\rm{b}} -
    \frac{3(\epsilon\gamma_3 k^2 + \gamma_4 H^2 a^2) }{4H a^2}
    \Omega_{\rm{b}} + \frac{3\epsilon r_{\rm{c}}\kappa_4^2 \rho a^3 \gamma_4 }{2k^2}
    \Delta;
\end{equation}
the following equation of motion for the density contrast of the
cold dark matter, $\Delta$:
\begin{equation}\label{eq:scalar brane equation}
    \ddot \Delta + 2H \dot \Delta - \frac{1}{2} \kappa_4^2 \rho
    \gamma_2 \Delta = - \frac{\epsilon \gamma_4 k^4}{4a^5}
    \Omega_{\rm{b}};
\end{equation}
and the following expressions for the two metric potentials $\Phi$
and $\Psi$:
\begin{align}
    \Phi & = +\frac{\kappa_4^2 \rho a^2 \gamma_1}{2k^2}\Delta +
    \frac{\epsilon \gamma_1}{4ar_{\rm{c}}} \dot\Omega_{\rm{b}} -
    \frac{\epsilon(k^2+3H^2a^2)\gamma_1}{12 Hr_{\rm{c}} a^3} \Omega_{\rm{b}}, \\
    \Psi & = -\frac{\kappa_4^2 \rho a^2 \gamma_2}{2k^2}\Delta +
    \frac{\epsilon \gamma_1}{4Hr_{\rm{c}} a } \ddot \Omega_{\rm{b}} -
    \frac{3\epsilon H \gamma_4}{4a} \dot\Omega_{\rm{b}} + \frac{\epsilon(k^2 r_{\rm{c}} \gamma_4 + Ha^2 \gamma_2)}{4r_{\rm{c}}
    a^3}\Omega_{\rm{b}}; \label{eq:Phi and Psi formulae}
\end{align}
where $\Omega_{\rm{b}} = \Omega_{\rm{b}}(t) =
\Omega(u_{\rm{b}}(t),v_{\rm{b}}(t))$ and $(\partial_y
\Omega)_{\rm{b}}$ are the values of the bulk field and its normal
derivative, respectively, evaluated at the brane. In these
expressions, the dimensionless $\gamma$-factors are functions of $H$
and $\dot{H}$.

The bulk wave equation (\ref{eq:scalar master eqn}), boundary
condition (\ref{eq:scalar boundary condition}) and (\ref{eq:scalar
brane equation}) are the equations we must solve. Once we know
$\Delta$ and $\Omega$ the metric perturbations $\Phi$ and $\Psi$ can
be obtained by differentiation.

\section{Scalar perturbations in the self-accelerating
universe}\label{sec:scalar SA}

In this section, we concentrate on the $w = \sigma = 0$ and
$\epsilon = +1$ DGP scenario as a model for the late-time
accelerating universe. By examining probes of the expansion
history,~\cite{Lazkoz:2007zk} it has been found that
\begin{equation}\label{eq:SA observational constraint}
    \Omega_{r_{\rm{c}}} = \frac{1}{4H_0^2 r_{\rm{c}}^2} = 0.15 \pm 0.02,
\end{equation}
at 95\% confidence. We use the best fit value for
$\Omega_{r_{\rm{c}}}$ in our simulations.

For all plots in this paper, we select the bulk field to be zero and
the brane field non-zero initially. We have also simulated several
different choices of initial data, such as the bulk field being
constant along the initial null hypersurface, and have found that
the simulation results remain the same as long as the initial time
is early enough. This is analogous to what happens in the RS
case.~\cite{Cardoso:2007zh}

In a previous work,~\cite{Koyama:2005kd} a `quasi-static' (QS)
approximation was developed to describe the behaviour of DGP
perturbations whilst well inside the cosmological horizon, $k \gg
Ha$, and with physical wavelengths much less than the crossover
scale, $a \ll kr_{\rm{c}}$. Here, we compare the QS approximation to
our simulations to determine just how large $k$ must be for it to be
valid.

\begin{figure}
\begin{center}
\includegraphics[height=10.8cm]{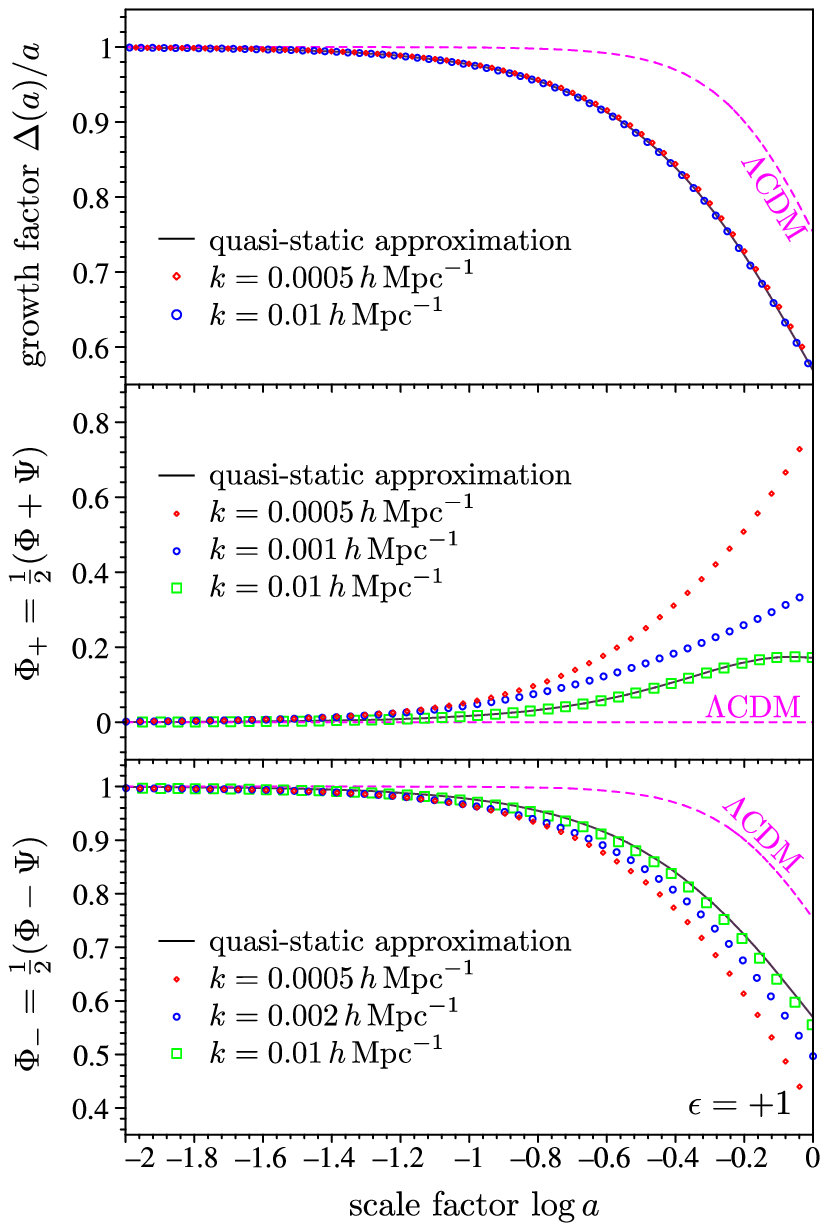}
\includegraphics[height=10.8cm]{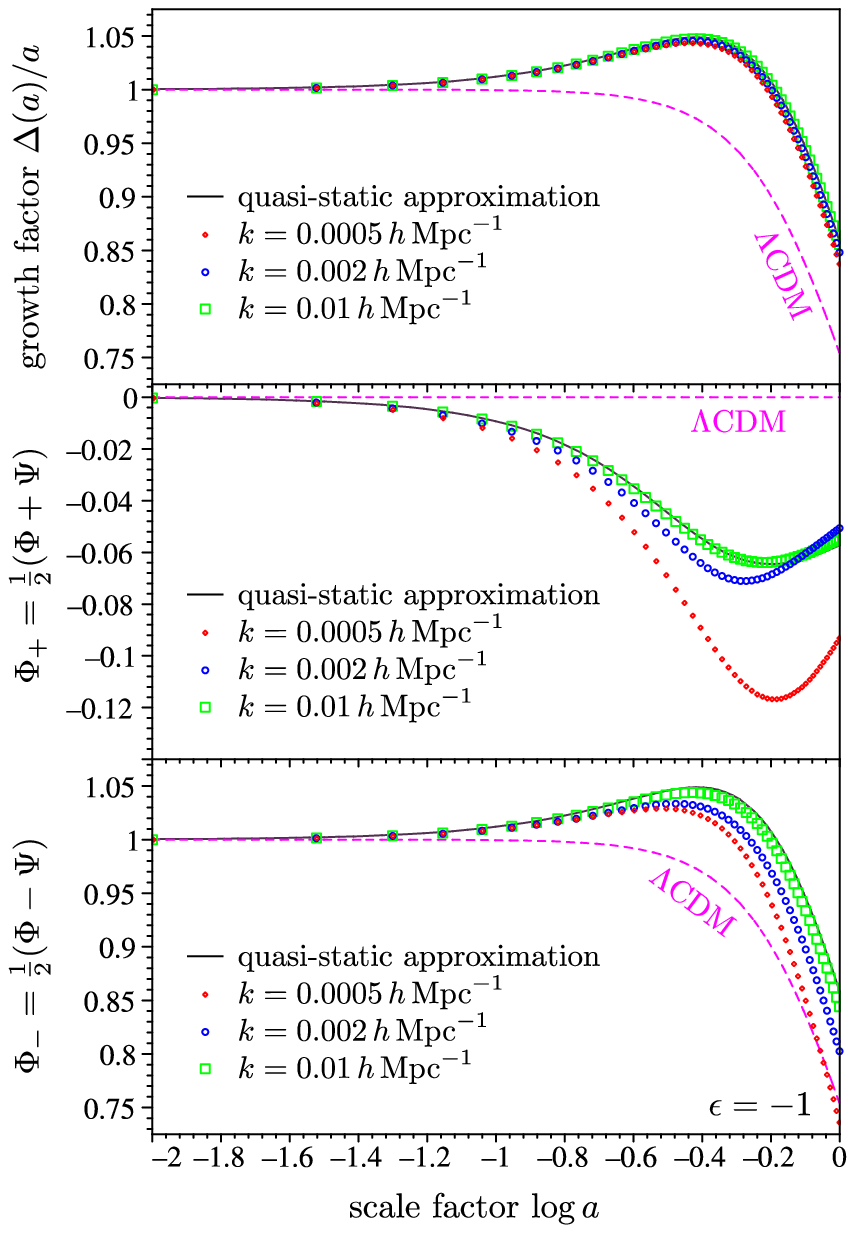}
\end{center}
\caption{Linear growth factor and alternate gravitational potentials
$\Phi_\pm$ from simulations and the QS approximation in the
self-accelerating branch (left) and normal branch with
$\Omega_{r_{\rm{c}}} = 0.05$ (right). We normalize $g(a)$ and
$\Phi_-$ to unity at early times. For comparison, we also show the
relevant results for the concordance $\Lambda$CDM model with
${\Omega_m} = 0.26$ and $\Omega_\Lambda =
0.74$.}\label{fig:quasistatic compare}
\end{figure}

In Fig.~\ref{fig:quasistatic compare} (left), we compare simulation
results versus the QS approximation for the linear growth factor
$g(a) = \Delta(a)/a$ and the alternate gravitational potentials
$\Phi_\pm = \tfrac{1}{2}(\Phi \pm \Psi)$. We see that the simulation
results are consistent with the QS approximation for $k \gtrsim
10^{-2}\,h\,\text{Mpc}^{-1}$. On larger scales, the potential
$\Phi_-$, which determines the integrated Sach-Wolfe (ISW) effect,
shows more suppression than the QS prediction.

\section{Scalar perturbations in the DGP normal
branch}\label{sec:scalar NB}

We now turn our attention to the behaviour of density perturbations
in the normal branch of the DGP model. Unlike the $\epsilon = +1$
case, this branch does not naturally have a late time accelerating
phase. So, in order to be made consistent with observations, we must
allow for the brane to have a nonzero tension that acts as an
effective 4-dimensional cosmological constant (we call this the
$\Lambda$DGP model). Assuming that the matter sector is
CDM-dominated, the Friedmann equation for this scenario follows from
the general form (\ref{eq:background dynamics}) with $\epsilon = -1$
and $w=0$. The background dynamics has been compared to observations
of $H(z)$,~\cite{Lazkoz:2007zk} and the following parameter values
were found:
\begin{equation}\label{eq:normal branch observational constraint}
    \Omega_\text{m} = \frac{\kappa_4^2 \rho_0}{3H_0^2} = 0.23\pm 0.04, \quad
    \Omega_{r_{\rm{c}}} = \frac{1}{4H_0^2 r_{\rm{c}}^2} \le 0.05,
\end{equation}
at $95\%$ confidence. Here, $\rho_0$ is the present day CDM density.
Note that the observationally preferred value of
$\Omega_{r_{\rm{c}}}$ is zero. Since the DGP model goes over to GR
in this limit, this implies that $\Lambda$CDM gives a better fit to
the data than $\Lambda$DGP. We will assume the best fit value of
0.23 for $\Omega_\text{m}$.

In Fig.~\ref{fig:quasistatic compare} (right), we compare the
results of our simulations to the QS approximation and $\Lambda$CDM
in the case $\Omega_{r_{\rm{c}}} = 0.05$. As in \S\ref{sec:scalar
SA}, we find that the simulation results are fairly insensitive to
initial conditions provided that the initial data surface is set far
enough into the past. In contrast to the self-accelerating case, we
find that the linear growth factor and $\Phi_-$ potential are
generally larger than in the $\Lambda$CDM case. The general trend is
for $\Phi_-$ to become larger on small scales. We also notice that
the QS approximation seems to provide a very good match to the
simulation results for $\Delta$ on all scales. Finally, as in the
self-accelerating case, we see that the QS approximation provides
reasonably accurate results (with errors $\lesssim 5\%$) on scales
$k \gtrsim 0.01\, h\, \text{Mpc}^{-1}$.

\section*{Acknowledgments}

AC is supported by FCT (Portugal) PhD fellowship SFRH/BD/19853/2004.


\section*{References}

\begin{thebibliography}{99}

\bibitem{Cardoso:2007xc}
  A.~Cardoso, K.~Koyama, S.~S.~Seahra and F.~P.~Silva,
  arXiv:0711.2563 [astro-ph].

\bibitem{Dvali:2000rv}
  G.~R.~Dvali, G.~Gabadadze and M.~Porrati,
  Phys.\ Lett.\  B {\bf 484} (2000) 112
  [arXiv:hep-th/0002190].

\bibitem{Dvali:2000hr}
  G.~R.~Dvali, G.~Gabadadze and M.~Porrati,
  Phys.\ Lett.\  B {\bf 485} (2000) 208
  [arXiv:hep-th/0005016].

\bibitem{Deffayet:2000uy}
  C.~Deffayet,
  Phys.\ Lett.\  B {\bf 502} (2001) 199
  [arXiv:hep-th/0010186].

\bibitem{Koyama:2005kd}
  K.~Koyama and R.~Maartens,
  JCAP {\bf 0601} (2006) 016
  [arXiv:astro-ph/0511634].

\bibitem{Cardoso:2006nh}
  A.~Cardoso, K.~Koyama, A.~Mennim, S.~S.~Seahra and D.~Wands,
  Phys.\ Rev.\  D {\bf 75} (2007) 084002
  [arXiv:hep-th/0612202].

\bibitem{Cardoso:2007zh}
  A.~Cardoso, T.~Hiramatsu, K.~Koyama and S.~S.~Seahra,
  arXiv:0705.1685 [astro-ph].

\bibitem{Sahni:2002dx}
  V.~Sahni and Y.~Shtanov,
  JCAP {\bf 0311} (2003) 014
  [arXiv:astro-ph/0202346].

\bibitem{Lue:2004za}
  A.~Lue and G.~D.~Starkman,
  Phys.\ Rev.\  D {\bf 70} (2004) 101501
  [arXiv:astro-ph/0408246].

\bibitem{Charmousis:2006pn}
  C.~Charmousis, R.~Gregory, N.~Kaloper and A.~Padilla,
  JHEP {\bf 0610} (2006) 066
  [arXiv:hep-th/0604086].

\bibitem{Mukohyama:2000ui}
  S.~Mukohyama,
  Phys.\ Rev.\  D {\bf 62} (2000) 084015
  [arXiv:hep-th/0004067].

\bibitem{Lazkoz:2007zk}
  R.~Lazkoz and E.~Majerotto,
  JCAP {\bf 0707} (2007) 015
  [arXiv:0704.2606 [astro-ph]].

\end{thebibliography}
\end{document}